\RequirePackage{fix-cm}
\documentclass[smallextended]{svjour3}       % onecolumn (second format)
\smartqed  % flush right qed marks, e.g. at end of proof
\usepackage{graphicx}
\usepackage{mathptmx}      % use Times fonts if available on your TeX system
\usepackage{latexsym}
\usepackage{amssymb}
\usepackage{amsmath}
\usepackage{amsfonts}
\newcommand{\be}{\begin{eqnarray}}
\newcommand{\ee}{\end{eqnarray}}
\newcommand{\bra}[1]{\mbox{$\langle\, #1 \mid$}}

\newcommand{\ket}[1]{\mbox{$\mid #1\,\rangle$}}

\newcommand{\pro}[2]{\mbox{$\langle\, #1 \mid #2\,\rangle$}}
\newcommand{\expec}[1]{\mbox{$\langle\, #1\,\rangle$}}

\renewcommand{\d}{\mbox{${\rm d}$}} 
\newcommand{\lp}{\ell_{\rm p}}
\newcommand{\mpl}{m_{\rm p}}

\newcommand{\rh}{r_{\rm H}}
\newcommand{\Rh}{R_{\rm H}}
\newcommand{\psis}{{\psi}_{\rm S}}

\newcommand{\psih}{{\psi}_{\rm H}}
\newcommand{\en}{{\mathcal{E}}}

\begin{document}
\title{Horizon Quantum mechanics: spherically symmetric and rotating sources\thanks{This work was partly
supported by INFN grant FLAG.}
}
\author{Roberto~Casadio
\and
Andrea~Giugno 
\and
Andrea~Giusti
\and
Octavian~Micu
}
\institute{R.~Casadio
              \at
              Dipartimento di Fisica e Astronomia, Universit\`a di Bologna, via Irnerio~46, I-40126 Bologna, Italy
              \\
              I.N.F.N., Sezione di Bologna, via B.~Pichat~6/2, I-40127 Bologna, Italy
              \\
              Tel.: +39-051-2091112
              \\
              \email{casadio@bo.infn.it}           %  \\
           \and
           A.~Giugno
           \at
           Arnold Sommerfeld Center, Ludwig-Maximilians-Universit\"at, Theresienstra{\ss}e 37, 80333 M\"unchen, Germany
           \and
           A.~Giusti
           \at
           Dipartimento di Fisica e Astronomia, Universit\`a di Bologna, via Irnerio~46, I-40126 Bologna, Italy
           \\
           I.N.F.N., Sezione di Bologna, via B.~Pichat~6/2, I-40127 Bologna, Italy
           \\
           Arnold Sommerfeld Center, Ludwig-Maximilians-Universit\"at, Theresienstra{\ss}e 37, 80333 M\"unchen, Germany
           \and
           O.~Micu
           \at
           Institute of Space Science, P.O.~Box MG-23, RO-077125 Bucharest-Magurele, Romania
}
\date{Received: date / Accepted: date}
\maketitle
\begin{abstract}
The Horizon Quantum Mechanics is an approach that allows one to analyse the gravitational radius
of spherically symmetric systems and compute the probability that a given quantum state is a black hole.
We first review the (global) formalism and show how it reproduces a gravitationally inspired GUP relation.
This results leads to unacceptably large fluctuations in the horizon size of astrophysical black holes
if one insists in describing them as (smeared) central singularities. 
On the other hand, if they are extended systems, like in the corpuscular models, no such issue arises
and one can in fact extend the formalism to include asymptotic mass and angular momentum with
the harmonic model of rotating corpuscular black holes.
The Horizon Quantum Mechanics then shows that, in simple configurations, the appearance of
the inner horizon is suppressed and extremal (macroscopic) geometries seem disfavoured.
\end{abstract}
\section{Introduction and motivation}
\setcounter{equation}{0}
The world as we know it, is best described by quantum physics, and black holes should reasonably
be represented as quantum objects as well.
A first non-trivial question that follows is how much of the classical description of black holes given by
general relativity we can still keep at the quantum level.
One feature that should presumably live up into the quantum realm is that black holes are ``gravitational
bound states''.
If so, another question one would like to answer is whether we can understand such states by
means of scattering theory and experiments.
Indeed, most of the existing literature simply analyses quantum effects on classical black hole
background space-times, and the proper quantum nature of the background itself is not fully
accounted for in this manner.   
Finally, one might notice that a classical quantity that characterises such bound states (in the particular
case when spherical symmetry is preserved) is the so called ``gravitational radius''.
It therefore appears that a quantum treatment of the gravitational radius could be a good way to start the
process of developing a fully quantum theory of black holes. 
\par
In fact, the Horizon Quantum Mechanics (HQM)~\cite{fuzzyh,hqft,gupf,acmo,Casadio:2015rwa,Casadio:2015qaq}
was precisely proposed with the purpose of describing the gravitational radius of spherically
symmetric compact sources and determining the existence of a horizon in a quantum mechanical fashion.
However, most astrophysical objects are observed to possess non-vanishing angular momentum, which led us
to extend the HQM to rotating sources~\cite{rotating}.
This is not a conceptually trivial task.
In a classical spherically symmetric system, the gravitational radius defined
in terms of the (quasi-)local Misner-Sharp mass, uniquely determines the location of the
trapping surfaces where the null geodesic expansion vanishes.
The latter surfaces are proper horizons in a time-independent configuration, which is
the case we shall always consider here.
It is therefore rather straightforward to uplift this description of the causal structure 
of space-time to the quantum level by simply imposing the relation between the
gravitational radius and the Misner-Sharp mass as a constraint to be satisfied
by the physical states of the system~\cite{hqft}.
In a non-spherical space-time, such as the one generated by an axially-symmetric rotating source,
although there are candidates for the quasi-local mass function that should replace the Misner-Sharp
mass~\cite{qlm-review}, the locations of trapping surfaces, and horizons, remain to be
determined separately.
In Ref.~\cite{rotating}, we therefore considered a different path and simply uplifted to a quantum
condition the classical relation of the two horizon radii with the mass
and angular momentum of the source obtained from the Kerr metric.
\par
Beside the formal developments, we applied the extended HQM 
to specific states with non-vanishing angular momentum of the harmonic black hole model
introduced in Ref.~\cite{QHBH} (see also Ref.~\cite{muck} for an improved version).
This model can be considered as a working realisation of the corpuscular black holes
proposed by Dvali and Gomez~\cite{dvali}, and turns out to be simple enough, so as to allow
one to determine explicitly the probability that the chosen states are indeed black holes.
We shall use units with $c=1$, and the Newton constant $G=\lp/\mpl$, where $\lp$ and $\mpl$
are the Planck length and mass, respectively, and $\hbar=\lp\,\mpl$.
\section{Horizon quantum mechanics}
\label{defH}
\setcounter{equation}{0}
We first review the basics of the global HQM for static spherically symmetric
sources~\cite{hqft}, and then extend this formalism to rotating systems.
\subsection{Spherically symmetric systems}
\label{sphS}
The general spherically symmetric metric $g_{\mu\nu}$ can be written as
\be
\d s^2
=
g_{ij}\,\d x^i\,\d x^j
+
r^2(x^i)\left(\d\theta^2+\sin^2\theta\,\d\phi^2\right)
\ ,
\label{metric}
\ee
where $r$ is the areal coordinate and $x^i=(x^1,x^2)$ are coordinates
on surfaces of constant angles $\theta$ and $\phi$.
In particular, if we set $x^1=t$ and $x^2=r$, and denote the static matter density as
$\rho=\rho(r)$, the Einstein field equations imply
\be
g^{rr}
=
1-\frac{2\,\lp\,(m/\mpl)}{r}
\ ,
\label{einstein}
\ee
where 
\be
m(r)
=
4\,\pi\int_0^r \rho(\bar r)\,\bar r^2\,\d \bar r
\ ,
\label{M}
\ee
is the Misner-Sharp mass.
A trapping surface exists if there are values of $r$ such that
the gravitational radius $\rh(r) = 2\,\lp\,{m(r)}/{\mpl}\ge r$.
In the vacuum outside the source,
$\rh$ becomes the usual Schwarzschild radius associated with the total 
Arnowitt-Deser-Misner (ADM)~\cite{adm} mass $M=m(\infty)$, 
\be
\Rh
=
2\,\lp\,\frac{M}{\mpl}
\ ,
\label{RhM}
\ee
which provides a mathematical foundation to
Thorne's {\em hoop conjecture\/}~\cite{Thorne:1972ji}.
\par
This description becomes questionable for sources of the Planck size or
lighter, since the Heisenberg principle introduces an uncertainty in the spatial localisation
of the order of the Compton-de~Broglie length $\lambda_M \simeq \lp\,{\mpl}/{M}$.
We could hence argue that $\Rh$ only makes sense if $\Rh\gtrsim \lambda_M$ ($M \gtrsim \mpl$).
The HQM assumes the existence of two observables, the quantum Hamiltonian
corresponding to the total energy $M$ of the system,
\be
\hat H
=
\sum_\alpha
E_\alpha\ket{E_\alpha}\bra{E_\alpha}
\ ,
\label{HM}
\ee
where the sum is over the Hamiltonian eigenmodes, and the gravitational radius
with eigenstates
\be
\hat R_{\rm H}\,\ket{{\Rh}_\beta}
=
{\Rh}_\beta\,\ket{{\Rh}_\beta}
\ .
\ee
General states for our system can correspondingly be described by linear combinations of 
the form
\be
\ket{\Psi}
=
\sum_{\alpha,\beta}
C(E_\alpha,{\Rh}_\beta)\,
\ket{E_\alpha}
\ket{{\Rh}_\beta} 
\ ,
%\label{Erh}
\ee
but only those for which the relation~\eqref{RhM} between the Hamiltonian and gravitational
radius holds are viewed as physical, that is
\be
0
=
\left(
\hat H
-
\frac{\mpl}{2\,\lp}\,\hat R_{\rm H}
\right)
\ket{\Psi}
=
\sum_{\alpha,\beta}
\left(
E_\alpha-\frac{\mpl}{2\,\lp}\,{\Rh}_\beta
\right)
C(E_\alpha,{\Rh}_\beta)\,
\ket{E_\alpha}
\ket{{\Rh}_\beta} 
\ .
\label{Hcond}
\ee
The solution of this contraint is given by
\be
C(E_\alpha,{\Rh}_\beta)
=
C(E_\alpha,{2\,\lp}\,E_\alpha/{\mpl})\,\delta_{\alpha\beta}
\ ,
\label{solG}
\ee
so that Hamiltonian eigenmodes and gravitational radius eigenmodes can only appear
suitably paired in a physical state.
In other words, the gravitational radius is not an independent degree of freedom in our treatment
(for a comparison with different approaches to the horizon quantisation, see section~2.4 in
Ref.~\cite{Casadio:2015qaq}).
\par
\begin{figure}[t]
\centering
\raisebox{5cm}{$P_{\rm BH}$}
\includegraphics[width=8cm]{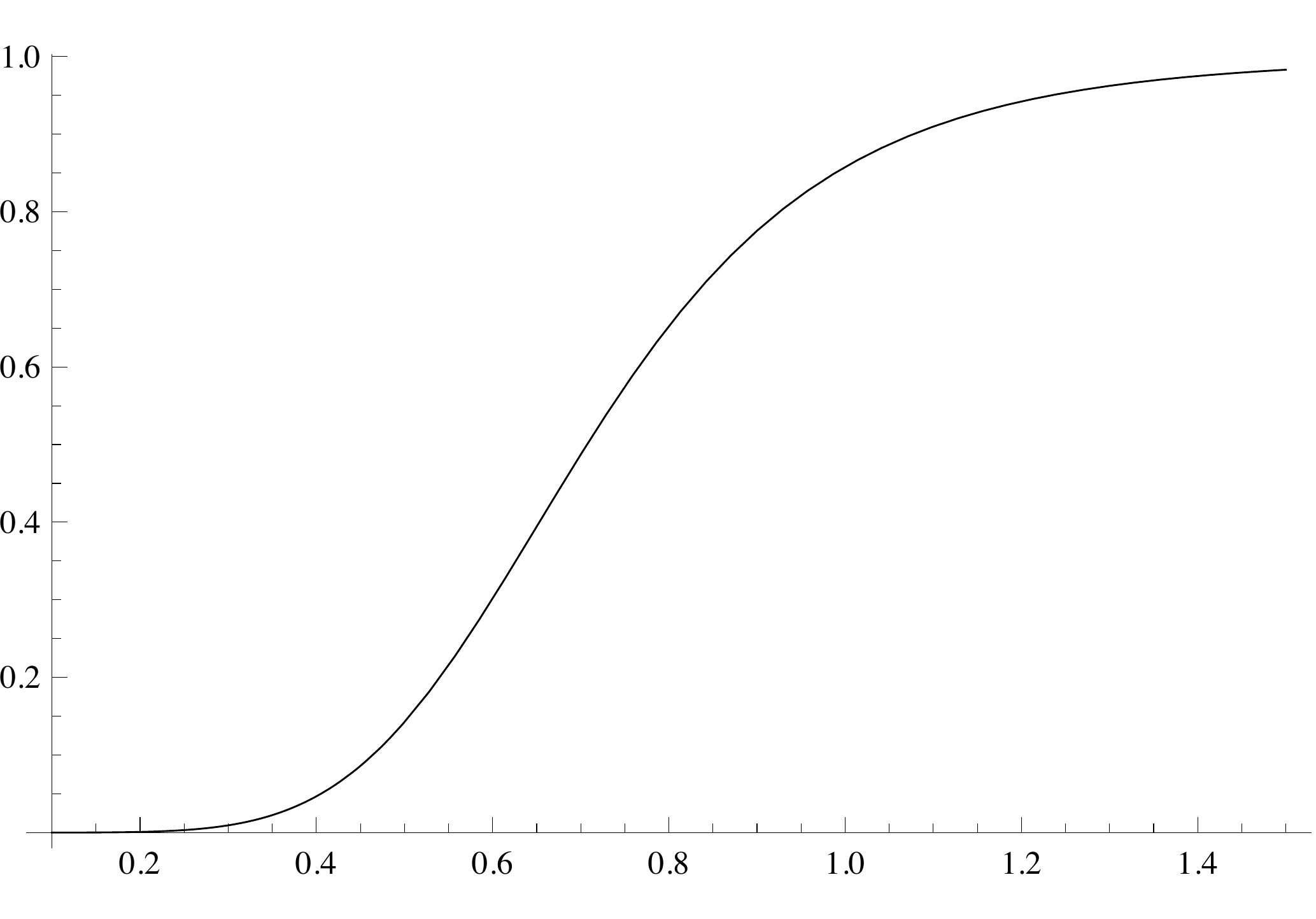}
\raisebox{0.5cm}{$m$}
\caption{Probability a Gaussian state of width $\ell=\hbar/m$ is a black hole (mass in Planck units).}
\label{pbhm}
\end{figure}

By tracing out the gravitational radius, we recover the spectral decomposition
of the source wave-function,
\be
\ket{\psis}
&=&
\sum_\gamma\bra{{\Rh}_\gamma}\,
\sum_{\alpha,\beta}
\ket{{\Rh}_\beta}\, 
C(E_\alpha,{2\,\lp}\,E_\alpha/{\mpl})\,\delta_{\alpha\beta}
\ket{E_\alpha}
\nonumber
\\
&\equiv&
\sum_\alpha
C_{\rm S}(E_\alpha)\,\ket{E_\alpha}
\ ,
\ee
where $C_{\rm S}(E_\alpha)\equiv C\left(E_\alpha,{2\,\lp}\,E_\alpha/{\mpl}\right)$
and we used the (generalised) orthonormality of the gravitational radius eigenmodes~\cite{hqft}.
Note that $\ket{\psis}$ is still a pure quantum state.
The Horizon Wave-Function (HWF) is instead obtained by integrating out the energy
eigenstates~\cite{fuzzyh,hqft}
\be
\psih({\Rh}_\sigma)
&=&
\pro{{\Rh}_\sigma}{\psih}
=
\bra{{\Rh}_\sigma}\,\sum_\gamma\bra{E_\gamma}\,
\sum_{\alpha,\beta}
\ket{E_\alpha}\, 
C(E_\alpha,{2\,\lp}\,E_\alpha/{\mpl})\,\delta_{\alpha\beta}
\ket{{\Rh}_\beta}
\nonumber
\\
&=&
C_{\rm S}({\mpl\,{\Rh}_\sigma}/{2\,\lp})
\ ,
\label{psihd}
\ee
where ${\mpl\,{\Rh}_\sigma}/{2\,\lp}=E({\Rh}_\sigma)$ is fixed by the constraint~\eqref{RhM}.
If the index $\sigma$ is continuous~\cite{hqft}, the probability density that we detect a
gravitational radius of size $\Rh$ associated with the quantum state
$\ket{\psis}$ is given by $\mathcal{P}_{\rm H}(\Rh) = 4\,\pi\,\Rh^2\,|\psih(\Rh)|^2$.
We can then define the conditional probability density that the source lies
inside its own gravitational radius $\Rh$ as
\be
\mathcal{P}_<(r<\Rh)
=
P_{\rm S}(r<\Rh)\,\mathcal{P}_{\rm H}(\Rh)
\ ,
\label{PrlessH}
\ee
where $P_{\rm S}(r<\Rh)= 4\,\pi \int_0^{\Rh} |\psis(r)|^2\,r^2\,\d r$.
Finally, the probability that the system in the state $\ket{\psis}$ is a black hole
is given by
\be
P_{\rm BH}
=
\int_0^\infty
\mathcal{P}_<(r<\Rh)\,\d \Rh
\ ,
\label{pbhgen}
\ee
which is plotted in Fig.~\ref{pbhm} for a Gaussian state of width $\ell$.
\par
\begin{figure}[t]
\centering
\raisebox{5cm}{$\Delta x$}
\includegraphics[width=8cm]{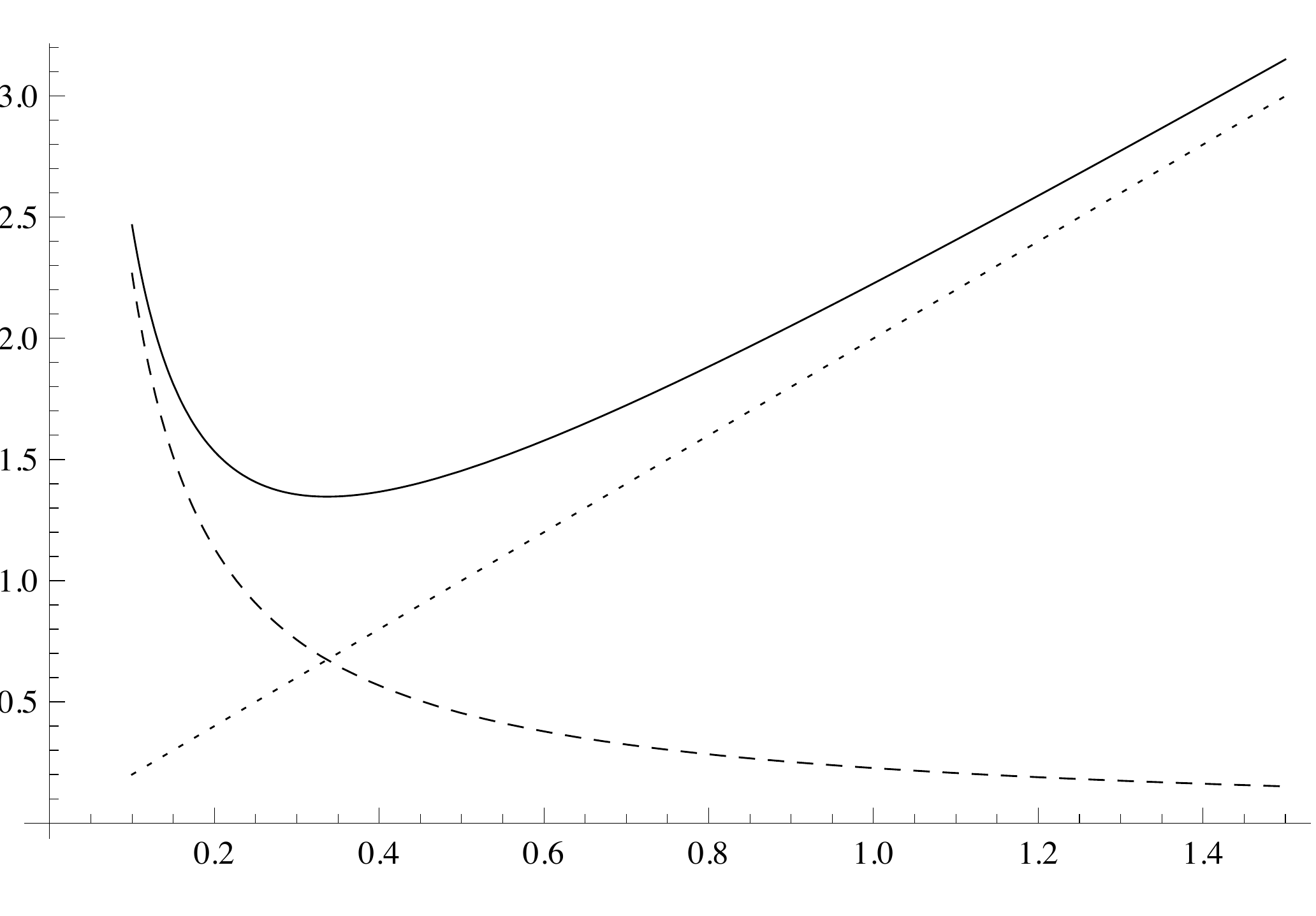}
\raisebox{0.5cm}{$\Delta p$}
\caption{GUP for a Gaussian source of width $\ell$ (all scales in Planck units).}
\label{gup}
\end{figure}

Note that now the gravitational radius is necessarily ``fuzzy'' and characterised by
an uncertainty $\Delta\Rh = \sqrt{\expec{\Rh^2}-\expec{\Rh}^2}$.
In fact, one can combine linearly the uncertainty $\Delta\Rh$ in the size of the gravitational radius 
with the analogue uncertainty in the source position $\Delta r$ and obtain a Generalised
Uncertainty Principle~\cite{gupf}
\be
\Delta x
&\equiv&
\sqrt{\expec{\Delta r^2}}
+
\sqrt{\expec{\Delta \rh^2}}
\nonumber
\\
&=&
\left(\frac{3\,\pi-8}{2\,\pi}\right)
\lp\,\frac{\mpl}{\Delta p}
+
2\,\lp\,\frac{\Delta p}{\mpl}
\ ,
\label{effGUP}
\ee
which is plotted in Fig.~\ref{gup} for Gaussian sources.
However, a serious issue immediately emerges if one insists in describing the source by
means of a sharply localised wave-function, such as a Gaussian of width $\ell$:
the uncertainty $\Delta\Rh\sim \ell^{-1}\sim \expec{\hat H}=M$, which means that 
astrophysical black holes would have wildly fluctuating horizons.
This is unacceptable and leads one to naturally consider models of black holes
with a source wave-function spread over large volumes of space, like the corpuscular
models of Refs.~\cite{dvali} or the harmonic models of Refs.~\cite{QHBH,muck}.
\subsection{Rotating systems}
\label{rotS}
In order to study rotating sources, since no analogue of the Misner-Sharp mass that can be used
to locate the horizons is known, we decided to exploit the asymptotic Kerr relations~\cite{rotating}.
We assume the existence of a complete set of commuting operators
$\{\widehat H, \, \widehat{J} ^{2}, \, \widehat{J} _{z} \}$ acting on a Hilbert space
$\mathcal H$ of the source. 
We also consider only the discrete part of the energy spectrum~\cite{hqft}, and
denote with $\alpha=\{a, \, j, \, m\}$ the set of relevant quantum numbers.
The source wave-function can thus be written as
\be
\ket{\psi _{S}} = \sum _{a, j, m} C_{S} (E_{a \, j}, \lambda _{j} , \xi _{m}) \, \ket{a \, j \, m}
\ ,
\label{eq-source-kerr}
\ee
where the sum formally represents the spectral decomposition in terms of the common
eigenmodes of the operators
\be
\hat H
&=&
\sum _{a, j , m} E_{a \, j} \, \ket{a \, j \, m} \bra{a \, j \, m}
\ ,
\\
\hat{J} ^{2} 
&=&
\mpl^{4} \, \sum _{a, j , m} j \,(j+1) \, \ket{a \, j \, m} \bra{a \, j \, m} \equiv \sum _{a, j , m} \lambda _{j} \, \ket{a \, j \, m} \bra{a \, j \, m}
\ ,
\\
\hat{J} _{z} 
&=&
\mpl^2 \,\sum _{a, j , m}  m \, \ket{a \, j \, m} \bra{a \, j \, m} \equiv \sum _{a, j , m} \xi _{m} \, \ket{a \, j \, m} \bra{a \, j \, m}
\ .
\ee
The quantum numbers $j \in \mathbb{N} _{0} /2$, $m \in \mathbb{Z} / 2$, with $|m| \leq j$,
and $a \in \mathcal{I}$, where $\mathcal{I}$ is a \textit{discrete}
set of labels that can be either finite of infinite.
\par 
For a stationary asymptotically flat space-time, we can still define the ADM mass
$M$ and replace this classical quantity with the expectation value of the Hamiltonian,
\be
M
\to
\bra{\psis} \hat H \ket{\psis}
&=&
\sum _{a, j, m}
\sum _{b, k, n}
C^{\ast} _{S} (E_{a \, j}, \lambda _{j} , \xi _{m}) \,
C_{S} (E_{b \, k}, \lambda _{k} , \xi _{n}) \, 
\bra{a \, j \, m} \widehat H \ket{b \, k \, n}
\notag
\\
&=& 
\sum _{a, j, m}
|C _{S} (E_{a \, j}, \lambda _{j} , \xi _{m})|^{2} \, E_{a \, j}
\ .
\ee
We can also employ the classical total angular momentum $J$ of the Kerr spacetime,
and lift it to a quantum observable~\cite{rotating},
\be
J^{2} \to \bra{\psis} \hat{J} ^{2} \ket{\psis}
=
\sum _{a, j, m} |C _{S} (E_{a \, j}, \lambda _{j} , \xi _{m})|^{2} \, \lambda_{j}
\ .
\ee 
We will further assume that $\expec{\hat{J}_{z}}$ is maximum in our quantum states,
so that the proper (semi-)classical limit is recovered,
\be 
\bra{\psis}\left( \hat{J} ^{2} - \hat{J} _{z}^2\right) \ket{\psis}
\
\underset{\hbar\to 0}{\overset{j \to \infty}{\longrightarrow}}
\
0
\ ,
\label{JJz}
\ee
for $\hbar\,j=\lp\,\mpl\,j$ held constant.
\par
In the Kerr space-time with $J^2<M^4$ there are two horizons given by
\be
\Rh^{(\pm)}
=
\frac{\lp}{\mpl}
\left(
M \pm \sqrt{M^{2} - \frac{J^{2}}{M^{2}}}
\right)
\ .
\label{R+-}
\ee
We correspondingly introduce two operators $\hat R ^{(\pm)}$ with eigenstates
\be
\hat R ^{(\pm)} _{\rm H}\,\ket{\beta} _{\pm}
=
{\Rh}^{(\pm)} _\beta\,\ket{\beta} _{\pm}
\ .
\ee
The generic state for our system can now be written as
\be
\ket{\Psi}
=
\sum_{a, j , m} \sum _{\alpha , \beta}
C(E_{a\, j}, \lambda _{j} , \xi _{m}, {\Rh} ^{(+)} _\alpha , {\Rh} ^{(-)} _\beta) \,
\ket{a \, j \, m}
\ket{\alpha}_{+}
\ket{\beta} _{-} 
\ ,
\label{Erh}
\ee
and, on defining the operators
\be
\hat{\mathcal{O}} ^{\pm}
\equiv
\hat H \pm \left( \hat{H} ^{2} - \hat{J} ^{2} \, \hat{H} ^{-2} \right) ^{1/2}
\, ,
\ee 
we obtain that the physical states must simultaneously satisfy
\be
\left(\hat R_{\rm H} ^{(\pm)} - \hat{\mathcal{O}} ^{\pm} \right)
\ket{\Psi} _{\rm phys}
=
0
\ .
\label{hR+-}
\ee
These two conditions reduce to
\be
C(E_{a\, j}, \{j \, m\}, {\Rh} ^{(+)} _\alpha , {\Rh} ^{(-)} _\beta)
&=&
C(E_{a\, j}, \{j \, m\}, {\Rh} ^{(+)} _{a\, j} (E _{a\, j}) , {\Rh} ^{(-)} _{a\, j} (E _{a\, j}))
\nonumber
\\
&&
\times
\delta _{\alpha , \{ a, j \}} \, \delta _{\beta , \{ a, j \}}
\ .
\ee	
By tracing out both radii states, we must recover the matter state~\eqref{eq-source-kerr},
which implies
\be
C_{S} (E_{a\, j}, \lambda_{j}, \xi_{m})
=
C(E_{a\, j}, \lambda_{j}, \xi_{m}, {\Rh} ^{(+)} _{a\, j} (E _{a\, j}) , {\Rh} ^{(-)} _{a\, j} (E _{a\, j}))
\ .
\ee 
By instead integrating away the matter states and only one of the two radii parts,
we can compute the wave-function corresponding to each horizon,
\be
\psi_\pm(\Rh^{(\pm)})
=
C(E _{a\, j}(\Rh^{(\pm)}), \lambda_{j}(\Rh^{(\pm)}), \xi_{m}(\Rh^{(\pm)}))
\ ,
\label{psi+-}
\ee
which imply a relation between the two horizons, $R_{H} ^{\pm} = R_{H} ^{\pm} (R_{H} ^{\mp})$.
\section{Corpuscular Harmonic Black Holes}
\setcounter{equation}{0}
\label{BECsection}
In the corpuscular model proposed by Dvali and Gomez~\cite{dvali}, black holes are macroscopic
quantum objects made of gravitons with a very large occupation number $N$ in the ground-state,
effectively forming Bose Einstein Condensates. 
These virtual gravitons forming the black hole of radius $\Rh$ can be effectively viewed
as being ``marginally bound'' by their Newtonian potential energy $U$, that is
\be
\mu + U_{\rm N}
\simeq
0
\ ,
\label{energy0}
\ee
where $\mu$ is the graviton effective mass related to their Compton size
$\lambda_\mu\simeq\lp\,{\mpl}/{\mu}\simeq\Rh$.
A first rough approximation for the binding potential energy is obtained by considering the 
square well expression
\be
U(r<\lambda_\mu)
\simeq
-N\,\alpha\,\mu
\ ,
\label{Udvali}
\ee
where the coupling constant $\alpha={\lp^2}/{\lambda_\mu^2}={\mu^2}/{\mpl^2}$.
Eq.~\eqref{energy0} then leads to $N\,\alpha = 1$ and
\be
M
\simeq
N\,\mu
\simeq
\sqrt{N}\,\mpl
\ .
\label{Max}
\ee
\subsection{Harmonic potential}
A better approximation for the potential energy is given by the harmonic form~\cite{QHBH}
\be
V
=
\frac{1}{2}\,\mu\,\omega^2\,(r^2-d^2)\,\Theta(d-r)
\equiv
V_0(r)\,\Theta(d-r)
\ ,
\label{Vho}
\ee
where the parameters $d$ and $\omega$ are chosen so as to ensure the highest energy
mode available to gravitons is marginally bound.
If we neglect the finite size of the well, the Schr\"odinger equation in spherical coordinates,
\be
\frac{\hbar^2}{2\,\mu\,r^2}
\left[
\frac{\partial}{\partial r} \left( r^2 \,\frac{\partial}{\partial r} \right)
+
\frac{1}{\sin \theta}\, \frac{\partial}{\partial \theta}
\left(  \sin \theta\, \frac{\partial}{\partial \theta} \right)
+ \frac{1}{\sin^2 \theta} \,\frac{\partial^2}{\partial \phi^2}
\right]
\psi
=
(V_0-\en)\,
\psi
\ ,
\label{schrod}
\ee
is solved by the well-known eigenfunctions
\be
\psi_{njm}(r,\theta,\phi;\lambda_\mu)
=
\mathcal{N}\,r^l\, e^{-\frac{r^2}{ 2\,\lambda_\mu^2}}\, _1F_1(-n,l+3/2,r^2/\lambda_\mu^2)\, Y_{lm}(\theta,\phi)
\ ,
\label{psin}
\ee
where $\mathcal{N}$ is a normalization constant, $_1F_1$ the Kummer confluent
hypergeometric function of the first kind and $Y_{lm}(\theta,\phi)$
are the usual spherical harmonics.
The corresponding eigenvalues are given by
\be
\en_{nl}
=
\hbar\, \omega
\left(2\,n+l+ \frac{3}{2}\right)
+V_0(0)
\ ,
\label{Enl}
\ee
where $n$ is the radial quantum number, and the quantum numbers $l$ and $m$ must not be confused
with the total angular momentum numbers $j$ and $m$ of Section~\ref{rotS}, as the latter are the sum
of the former.
Moreover, the eigenvalues $\en_{nl}$ must not be confused with the ADM energy $E_{aj}$, which
equals $N\,\mu$ by construction~\cite{baryons}.
\par
If we denote with $n_0$ and $l_0$ the quantum numbers of the highest eigenstate, 
and include the graviton effective mass $\mu$ in the constant $V_0(0)$,
the condition~\eqref{energy0} becomes $\en_{n_0 l_0}\simeq 0$, which yields
\be
\omega\,d^2
\simeq
2\,\frac{\hbar}{\mu}
\left(2\,n_0+ l_0 + \frac{3}{2}\right)
\ .
\ee
We now further assume that $d\simeq\lambda_\mu\simeq\Rh^{(+)}$ and use the Compton
relation for $\mu$, so that the potential can be finally written as
\be
V_0
=
2\,\mu
\left(2\,n_0+ l_0 + \frac{3}{2}\right)^2
\frac{r^2-\lambda_\mu^2}{\lambda_\mu^2}
\ee
and the eigenvalues as
\be
\en_{nl}
\simeq
-2\,\mu
\left(2\,n_0+ l_0 + \frac{3}{2}\right)
\left[2\,(n_0-n)+( l_0-l)\right]
\ ,
\ee
which of course holds only for $n\le n_0$ and $l\le l_0$.
The fact $\en_{nl}\le 0$ agrees with the post-Newtonian analysis of the
``maximal packing condition'' for the virtual gravitons in the black hole~\cite{baryons}.
\subsection{Rotating Black Holes}
In Ref.~\cite{rotating}, we considered in details some specific configurations of harmonic black holes with
angular momentum for which we applied the extended HQM described in the previous section.
For brevity, we shall here only recall the main results.
%
%We first remark that the quantum state of $N$ identical gravitons will be a $N$-particle
%state, i.e.~a vector of the $N$-particle Fock space $\mathcal{F} = \mathcal{H} ^{\otimes N}$,
%where $\mathcal{H}$ is a suitable 1-particle Hilbert space. 
%However, both the Hamiltonian of the system $\hat{H}$ and the gravitational radius
%$\hat R_{\rm H}$ are global observables and act as $N$-body operators on $\mathcal{F}$.
%
%
%\subsubsection{Single eigenstates}
%
%
\par
The simplest configuration corresponds to all toy gravitons in the same mode,
\be 
\ket{\Psi} \equiv \ket{M \, J} = \bigotimes _{\alpha = 1} ^N \ket{g} _\alpha
\ ,
\ee
where $\ket{g}$ represents the wave-function of a single component and the total ADM energy
is given by
\be
\bra{\Psi}\hat H\ket{\Psi}
\equiv
\expec{\hat H}
=
N\,\mu
=
M
\ ,
\ee
Each graviton is then set to have $n=n_0=0$, $l=l_0=2$ and $m=\pm 2$, that is
\be
\expec{r, \theta, \phi \, | \, g} =
\psi_{02\pm2} (r, \theta, \phi;\lambda_\mu)
=
\mathcal{N}\,r^2\,\exp\left(- \frac{r^2}{2\,\lambda_\mu^2} \right) \,Y_{2\pm2}(\theta,\phi)
\ ,
\label{psi022}
\ee
where the normalisation constant $ \mathcal{N} = {4}/({\sqrt{15} \, \pi ^{1/4} \, \lambda _\mu ^{7/2}})$.
The total angular momentum is thus given by 
\be
\bra{\Psi}\hat J^2\ket{\Psi}
\equiv
\expec{\hat{J} ^2}
&\!\!=\!\!&
4\,(N_+-N_-)(N_+-N_-+1/2)\,\mpl^2
\nonumber
\\
&\!\!\equiv\!\!&
4\,L^2\,N^2\,\mpl^2
\ ,
\ee
where $N_+\ge N_-=N-N_+$ is the number of spin up constituents (with $m=+2$).
Note that $L^2=0$ for $n_+\equiv N_+/N=1/2$ (the non-rotating case with $N_+=N_-$) and grows
to a maximum $L^2\simeq 1+\mathcal{O}(1/N)$ for the maximally rotating case $n_+=1$ (or $N_+=N$).
\par
Since we are considering an eigenstate of both the Hamiltonian $\hat H$ and total
angular momentum $\hat J^2$, the wave-functions~\eqref{psi+-} for the two horizons 
will reduce to single eigenstates of the respective gravitational radii, with
\be
\expec{\hat R_{\rm H}^{(\pm)}}
=
N\,\lp\,\frac{\mu}{\mpl}
\left(
1
\pm
\sqrt{1
-
4\,\frac{L^2\,\mpl^4}{N^2\,\mu^4}}
\right)
\ .
\label{expRh}
\ee
The classical condition for the existence of these horizons is that the square root be real,
which implies
\be
\mu^2
\ge
2\,\mpl^2\,\frac{L}{N}
\ .
\ee
This bound vanishes for $N_+=N_-=N/2$, as expected for a spherical black hole, 
and is maximum for $N_+=N$, in which case it yields
$\mu^2\gtrsim 2\,{\mpl^2}/{N}$, for $N\gg 1$.
\begin{figure}[t]
\centering
\includegraphics[width=8cm]{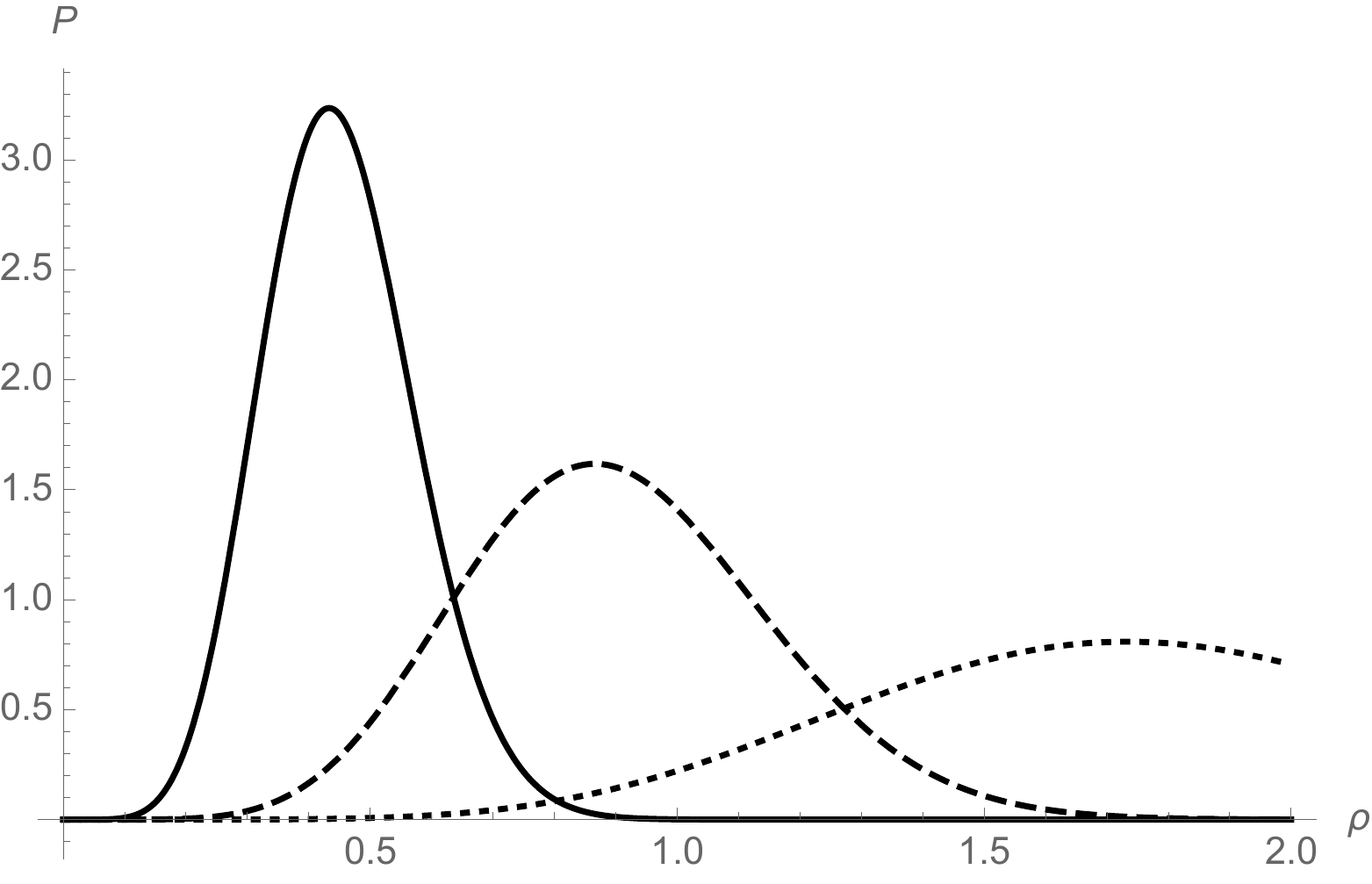}
\caption{Plots of the single-particle probability density as a function of
$\rho=r/\expec{\hat R_{\rm H}^{(+)}}$ for
$\lambda_\mu= \expec{\hat R_{\rm H}^{(+)}}/4$ (solid line),
$\lambda_\mu= \expec{\hat R_{\rm H}^{(+)}}/2$ (dashed line)
and $\lambda_\mu= \expec{\hat R_{\rm H}^{(+)}}$ (dotted line).}
\label{Pr}
\end{figure}
\par
\begin{figure}[t]
\centering
\includegraphics[width=8cm]{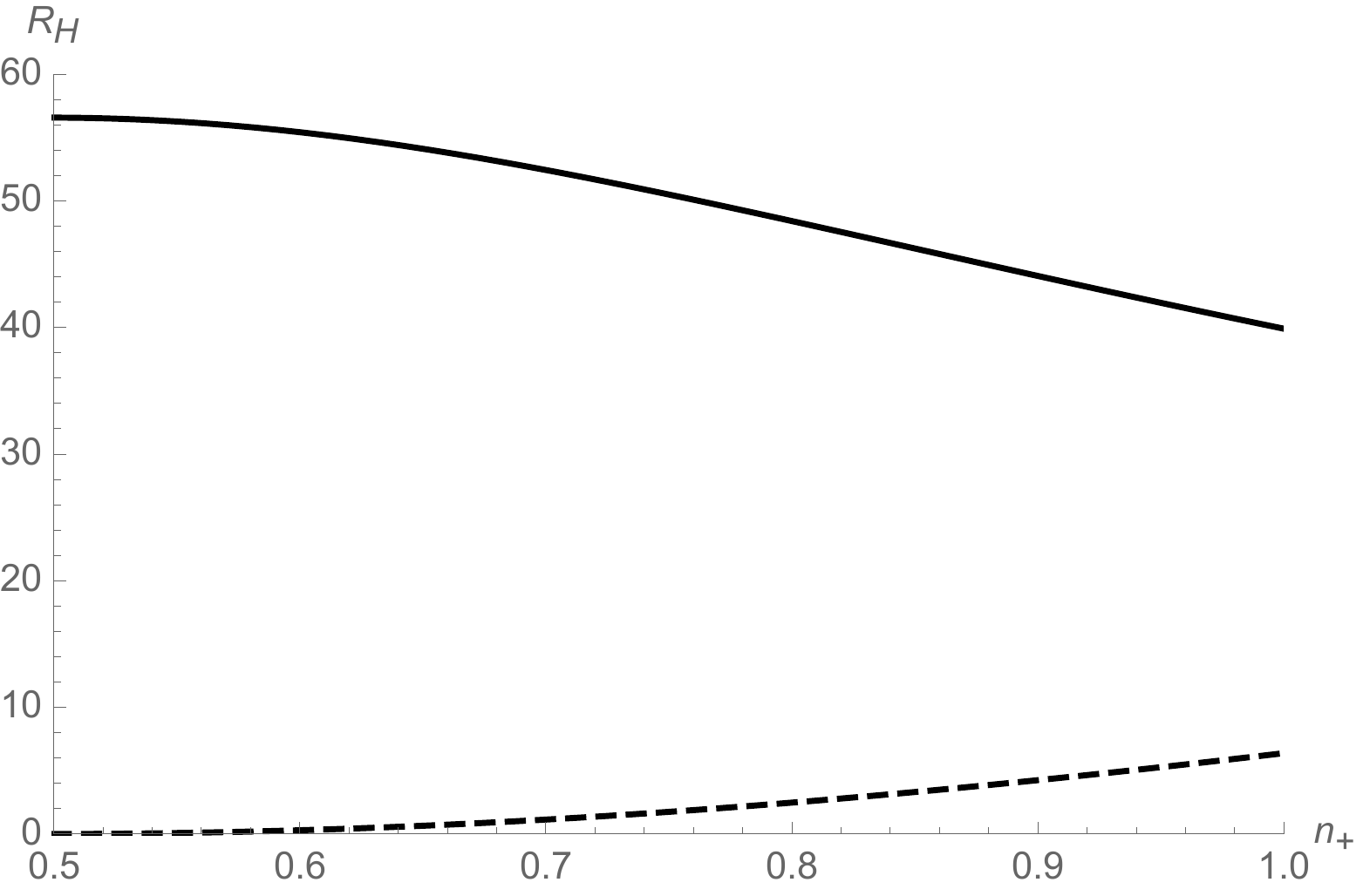}
\caption{Horizon radius $\expec{\hat R_{\rm H}^{(+)}}$ (solid line) and $\expec{\hat R_{\rm H}^{(-)}}$
(dashed line) in Planck length units for $N=100$ and $\gamma=2$.}
\label{RHpm2}
\end{figure}

Since we are modelling black holes, we next require that the Compton length $\lambda_\mu=\lp\,\mpl/\mu$,
is such that the modes~\eqref{psi022} are mostly found inside the outer horizon radius
$\expec{\hat R_{\rm H}^{(+)}}$.	
From the single-particle probability density displayed in Fig.~\ref{Pr}, we see that this occurs
for $\lambda_\mu=\expec{\hat R_{\rm H}^{(+)}}/4$,
whereas $\lambda_\mu=\expec{\hat R_{\rm H}^{(+)}}/2$ is already borderline
and $\lambda_\mu=\expec{\hat R_{\rm H}^{(+)}}$ is unacceptable.
We find it in general convenient to introduce the variable
\be
\gamma(n_+,N)
\equiv
\frac{\expec{\hat R_{\rm H}^{(+)}}}{2\,\lambda_\mu}
\ ,
\label{gamma}
\ee
and, after some calculations, the horizon radii for $N\gg 1$ and $\gamma\gtrsim 1$ are found to be given by 
\be
\expec{\hat R_{\rm H}^{(-)}}
\simeq
\frac{L^2}{\gamma^2}\,\expec{\hat R_{\rm H}^{(+)}}
\simeq
2\,L^2\,\lp\,\sqrt{\frac{N/\gamma}{\gamma^2+L^2}}
\ .
\label{1rh-}
\ee
Fig.~\ref{RHpm2} displays the result for $1/2\le n_+\le 1$ with $\gamma=2$ 
and $N=100$.
\par
\begin{figure}[t]
\centering
\includegraphics[width=8cm]{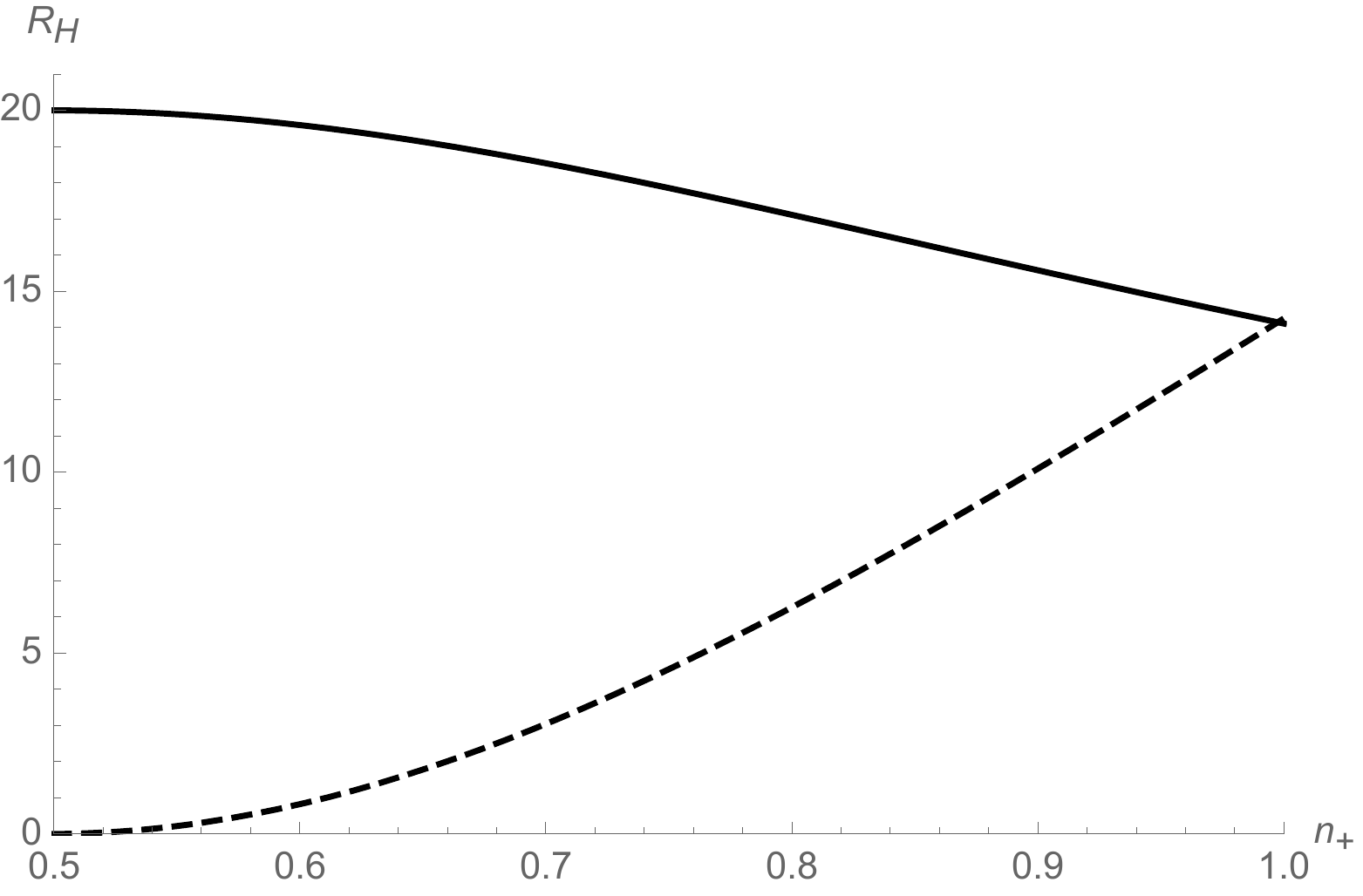}
\caption{Horizon radius $\expec{\hat R_{\rm H}^{(+)}}$ (solid line) and $\expec{\hat R_{\rm H}^{(-)}}$
(dashed line) in Planck length units for $N=100$ and $\gamma=1$.}
\label{RHpm}
\end{figure}

The extremal Kerr geometry can only be realised in our model if $\gamma$ is sufficiently small.
In fact, $\expec{\hat R_{\rm H}^{(-)}}\simeq\expec{\hat R_{\rm H}^{(+)}}$ requires
$\gamma^2\simeq L^2$.
For $\gamma=1$ and $N=100$, the horizon structure is displayed in Fig.~\ref{RHpm}, 
where we see that the two horizons meet at $L^2\simeq 1$, that is the configuration
with $n_+\simeq 1$ in which (almost) all constituents are aligned.
Note also that, technically, for $N\simeq 1$ and $\gamma$ small, there would be a finite range
$n_{\rm c}<n_+\le 1$ in which the expressions of the two horizon radii switch.
However, this result is clearly more dubious as one would be dealing with a truly quantum
black hole made of a few constituents just loosely confined.
\par
\begin{figure}[t]
\centering
\includegraphics[width=8cm]{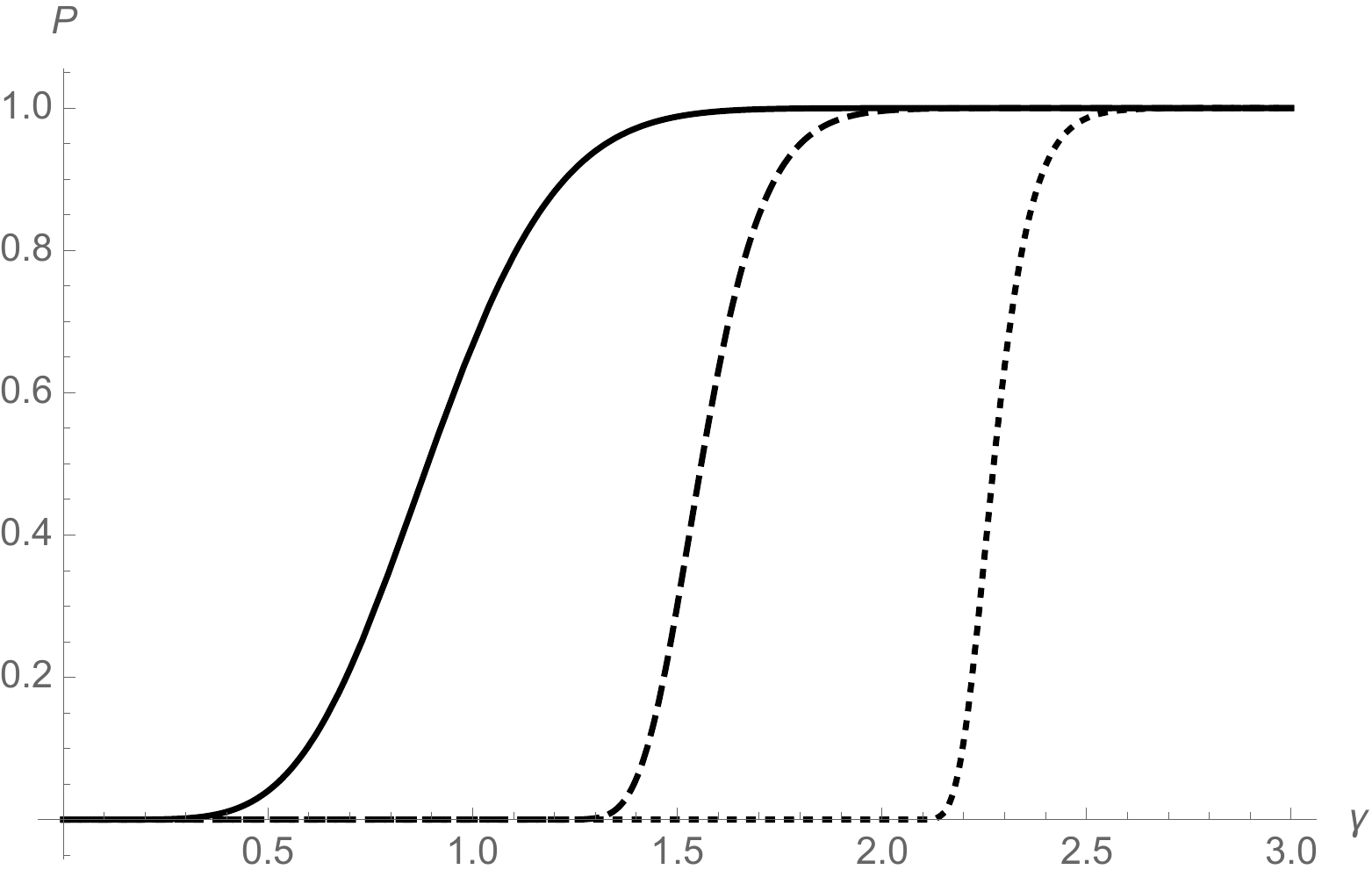}
\caption{Black hole probability~\eqref{1mpbh} as a function of $\gamma=\expec{\hat{R} ^{(+)} _{\rm H}}/2\,\lambda_\mu$
for $N=1$ (solid line), $N=10^2$ (dashed line) and $N=10^6$ (dotted line).}
\label{p<1}
\end{figure}

Finally, we can apply the HQM and compute the probability that the system discussed above is indeed
a black hole.
This is given by Eq.~\eqref{pbhgen} for the outer horizon, and turns out to be~\cite{rotating}
\be
P_{\rm BH} (n_+ , N) 
=
\prod _{\alpha = 1}^N
P_< ^{(+)} (r_\alpha < \expec{\hat{R} ^{(+)} _{\rm H}})
=
\left[P_< ^{(+)} (r < \expec{\hat{R} ^{(+)} _{\rm H}})\right]^N
\ ,
\label{1mpbh}
\ee
where
\be
P_< ^{(+)}(r< \expec{\hat{R} ^{(+)} _{\rm H}})
=
{\rm erf}\left(2\,\gamma\right)
-\frac{4}{15\,\sqrt{\pi}}\,\gamma
\left(
64\,\gamma^4
+40\,\gamma^2
+15
\right)
e^{-4\,\gamma^2}
\ ,
\label{1pbh}
\ee
is the single-particle ($N=1$) black hole probability, and we recall $\gamma$ depends on $N$ and $n_+$.
Eq.~\eqref{1pbh} is represented by the solid line in Fig.~\ref{p<1}, from which it is clear that it practically
saturates to 1 for $\gamma\gtrsim 2$. 
The same graph shows that the minimum value of $\gamma$ for which
$P_{\rm BH}(n_+,N)$ approaches 1 increases with $N$ (albeit very slowly~\cite{rotating}).
For $\gamma=1$, which we saw can realise the extremal Kerr geometry, we find 
\be
P_{\rm BH}(n_+,N)
\simeq
P_{\rm BH}(N)
\simeq
(0.67)^N
\ ,
\ee
and the system is most likely not a black hole for $N\gg 1$, in agreement with the probability density
shown in Fig.~\ref{Pr}.
One might indeed argue this probability is always too small for a (semi)classical black hole,
and the extremal Kerr configuration is therefore more difficult to achieve. 
\par
Analogously, we can compute the probability $P_{\rm IH}$ that the inner horizon is realised.
It is fairly obvious that, for any fixed value of $\gamma$, $P_{\rm IH}(n_+,N)\le P_{\rm BH}(n_+,N)$ 
and that equality is reached at the extremal geometry with
$\expec{\hat R_{\rm H}^{(-)}}\simeq \expec{\hat R_{\rm H}^{(+)}}$.
Moreover, from $0\le L^2\le 1$ and Eq.~\eqref{1rh-},
we find $\expec{\hat R_{\rm H}^{(-)}}\lesssim \expec{\hat R_{\rm H}^{(+)}}/\gamma^2$,
so that for $\gamma=2$, the probability $P_{\rm IH}\lesssim (0.04)^N$ is totally negligible for $N\gg 1$.
This suggest that the inner horizon can remain extremely unlikely even in configurations that should represent
large (semi-)classical black holes.
%
%
%
%\subsubsection{Superpositions}
%
\par
In Ref.~\cite{rotating}, we also investigated general superpositions of the states considered above,
but we shall not report on such cases here.
\section{Conclusions}
\label{conc}
\setcounter{equation}{0}
After a brief review of the original HQM for static spherically symmetric sources,
we have reported on its generalisation to rotating sources given in Ref.~\cite{rotating}.
This extension is not based on (quasi-)local quantities, but rather on the asymptotic mass and angular
momentum of the Kerr class of space-times.
The main results of our construction are that extremal configurations appear unlikely to have a
horizon, and that the inner horizon has a small probability to exist even for large sources.
\section*{Acknowledgments}
R.C.~and A.G.~are partially supported by the INFN grant FLAG.
The work of R.C.~and A.G.~has also been carried out in the framework of activities of the
National Group of Mathematical Physics (GNFM, INdAM).
O.M.~was supported by the grant LAPLAS~4. 
%
%%%%%

%
\end{document}